\documentclass[aps,prb,twocolumn,groupedaddress,showpacs]{revtex4}
\usepackage{amsmath}
\usepackage{dcolumn}
\usepackage{amssymb}
\usepackage{amsfonts}
\usepackage{graphicx}
\usepackage[T1]{fontenc}
\usepackage{indentfirst}
\usepackage{color}
\usepackage{bbm}

\begin{document}
\title{Machine learning quantum criticality in the spin-1/2 quantum antiferromagnets on the square lattice with plaquette structure}
\author{Tanja \DJ uri\'c}
\affiliation{Department of Physics, Faculty of Science, University of Zagreb, Bijeni\u{c}ka cesta 32, 10000 Zagreb, Croatia}

\date{\today}
\begin{abstract}
The power of machine learning algorithms to automatically classify different phases of matter and detect quantum phase transitions without necessity to characterize phases by various quantities like local order parameters or topological invariants as in conventional approaches defined machine learning phases of matter as a new research frontier and basic research tool in condensed matter and statistical physics. We study quantum criticality in the spin-$1/2$ square-lattice $J_1$-$J_2$ model with additional plaquette structure by combination of reinforcement and supervised machine learning techniques. In our calculations the ground-state spin-spin correlation matrices for several system sizes are first found by restricted Boltzmann machine based variational Monte Carlo method, equivalent to reinforcement learning, and then used as a training data for convolutional neural network based supervised machine learning algorithm for phases classification. The model exhibits a quantum phase transition from paramagnetic plaquette resonating valence bond state to an antiferromagnetic state and has been a topic of great interest because of its close connection to cuprate superconductors and possibility of realization in cold atoms experiments. We consider both frustrated  and unfrustrated regimes and compare our results with the results obtained previously with other methods.  In the presence of frustration quantum Monte Carlo calculations are inhibited by negative sign problem and the results from previous calculations are available only for the unfrustrated case. We find that our results are in good agreement with the results obtained by coupled cluster and real space renormalization group methods for both frustrated and unfrustrated regimes. The quantum Monte Carlo and finite-size scaling result for the unfrustrated case however slightly differs from our result for the critical value of the inter-plaquette coupling strength, although the results are still in reasonable agreement. 
\end{abstract} 


\maketitle
\section{Introduction}
\label{sec:Introduction}
Machine learning (ML) field of computer science that is core of artificial intelligence, in particular deep learning, showed to be a very powerful tool for many modern technologies tasks requiring feature learning and pattern recognition such as image classification , natural language processing, speech recognition , and video games development, with deep learning being particularly successful at uncovering features in structured data (feature learning and compression).\cite{Hershey,Silver,LeCun,Goodfellow,Kelleher} Success of deep learning algorithms can be explained by their connection to variational renormalization group (RG)\cite{Mehta, Chung, Koch,Janusz,Funai} which is an iterative coarse-graining scheme that allows extraction of relevant features (operators) as a physical system is studied at different length scales and one of the most important and successful techniques in theoretical physics.

Recently artificial neural networks (ANN) and ML techniques also proved to be very powerful methods for studying variety of complex many-body problems\cite{Mechta1,Carleo1,DasSarma,Jia,Carleo,Carleo2,Szabo,Shi,Borin,Wu,Choo,Vieijra,Nomura,Nomura2,Astrakhantsev,Irikura,Choo2,Westerhout,Hartmann,Schmitt,Vicentini,Nagy,Yoshioka,Liang,Gao,Deng,Torlai,Deng2,Saito,Sharir,Kaubruegger,Duric,Pilati,Deng3} and is therefore of great importance to further investigate applicability of these methods in studying challenging models for which exact solutions are not known. In this paper we study one of such models, spin-$1/2$ square lattice $J_1$-$J_2$ model with additional plaquette structure, and demonstrate suitability of ML techniques to study complex phases and phase transitions by comparing the ML results with the results obtained by other available methods.

ML algorithms are designed to deal with large and complex data sets and make predictions on data by classifying and extracting features from data and are therefore very successful in studying various complex quantum systems where complete description of quantum many-body states requires exponentially large data sets. Namely in ML algorithms machine is an ANN that can learn probability distribution over set of its inputs. These networks originally introduced as simplified models of human brain can also be used to construct compact representations of many-body quantum states where the many-body wave-function corresponds to the probability distribution that the network tries to approximate. 

Boltzmann machines, in particular restricted Boltzmann machines (RBMs), widely used in machine learning community, showed very promising results so far since RBMs can represent many quantum states of interest and can also be efficiently numerically optimized using variational Monte Carlo (VMC) method.\cite{Carleo} Additionally there are strong connections between restricted Boltzmann machines and some classes of tensor network states in arbitrary dimensions, for example Jastrow wave-functions, entangled plaquette states (EPS) and string-bond states (SBS).\cite{Clark,Glasser,Chen} In ML language VMC optimization of an ANN is equivalent to reinforcement learning (RL)\cite{Sutton, Otterlo} where an agent learns from an interactive environment by trial and error using feedback from its own actions and experiences. The agent takes a suitable action to maximize reward in a particular situation and creates an action-reward feedback loop of an RL algorithm. Mathematical frameworks to describe an environment in RL are Markov decision processes and within VMC method a wave-function is learned on the basis of feedback from variational principle.

ML techniques, both in supervised and unsupervised forms, also proved to be very powerful techniques for detecting phase transitions in various systems. \cite{Mechta1,Carleo1,Carrasquilla,Carrasquilla2,Lian,Iakovlev,Broecker,Dong,Hsu,Hu,Chng,Zhang,Wang,Nieuwenburg,Broecker2,Yoshioka2,Wetzel,RodriguezNieva,Venderley,Giannetti,Zhang2,Zhang3,Berezutskii,Kharkov,Rem,Rao,Canabarro,Wetzel2,Morningstar,Zhao1,Cheng,Jadrich,Kim,Bohrdt,Rzadkowski,Yao,Che,Ohtsuki,Zlabys,Tan} ML allows automatic classification of different phases of matter without necessity to characterize phases by local order parameters, topological invariants or many-body localized phases as in conventional approaches. Such power of ML techniques to extract information of phases and phase transitions directly from many-body configurations where classification within an ANN occurs without any knowledge of the Hamiltonian or locality of interactions defined ML phases of matter as a new research frontier and a basic research tool in the field of condensed matter and statistical physics to identify poorly understood phases where the order parameter or topological description are not known a priori. 
\begin{figure}[t!]
\includegraphics[width=\columnwidth]{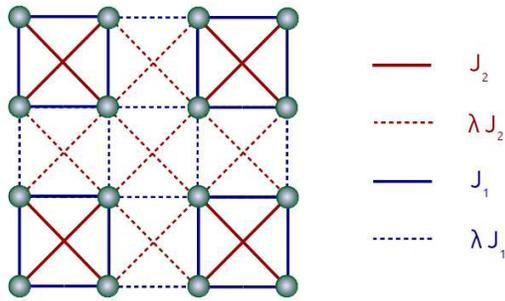}
\caption{\label{Fig:Hamiltonian} Visualization of the spin-$1/2$ two-dimensional square lattice Heisenberg model with plaquette structure (\ref{eq:Hamiltonian}). NN and NNN antiferromagnetic couplings correspond to lattice bonds illustrated with blue and red lines, respectively. Solid lines correspond to intra-plaquette couplings ($J_1$ and $J_2$) whilst dashed lines correspond to inter-plaquette couplings ($\lambda J_1$ and $\lambda J_2$). For $\lambda=1$ the model corresponds to the standard uniform spin-$1/2$ $J_1$-$J_2$ Heisenberg antiferromagnet with NN and NNN couplings, and for $\lambda=0$ the Hamiltonian describes unconnected four-spin $J_1$-$J_2$ plaquettes. At fixed $J_2/J_1$ a quantum phase transition from paramagnetic plaquette resonating valence bond state to a state with antiferromagnetic long-range order occurs at some critical value $\lambda_c$ of the inter-plaquette coupling strength parameter. 
}
\end{figure}

In an unsupervised ML algorithm only input to the algorithm is a set of data and the algorithm attempts to extract the features from the data or to arrange data into clusters.\cite{Mechta1} On the other hand, the basic idea of supervised machine learning (SML) is to train a machine (ANN) with large previously solved data set of input-output pairs and then use trained ANN model to process, characterize and make predictions for new data.\cite{Mechta1} Training procedure for an ANN is an optimization problem where a non-linear variational function is optimized with respect to a loss (cost) function by slowly adjusting the free parameters associated with connections between ANN neurons and their biases until high classification accuracy is obtained. Loss function is a function that gives information how good an ANN is for a certain task.

A typical example of SML is image recognition where a large set of labeled images is used as a training set, for example, images of cats (label $1$) and dogs (label $0$) with images as input and image names (labels) as output. After training procedure trained machine (ANN) that has learned key features of training set of labeled images is used to make predictions (perform a task), that is to recognize new images that are not in the training data set as being images of a cat or a dog. ANNs particularly successful in image recognition and classification are convolutional neural networks (CNNs), a class of deep neural networks that take into consideration additional input data set symmetries and structure (locality and translation invariance). Taking into account hierarchical pattern in data CNNs can construct more complex patterns by considering smaller and simpler patterns.

In this paper we study spin-$1/2$ square lattice $J_1$-$J_2$ model with plaquette structure using combination of RBM based RL (VMC) and CNN based SML methods. The model Hamiltonian is:
\begin{equation}\label{eq:Hamiltonian}
H=\sum_{a=0}^{1}(\delta_{0,a}+\delta_{1,a}\lambda)(J_1\sum_{\langle i,j \rangle_a}\vec{S}_i\vec{S}_j + J_2\sum_{\langle\langle i,j\rangle\rangle_a}\vec{S}_i\vec{S}_j)
\end{equation}
where $\vec{S}_i$ are spin-$1/2$ operators, $\delta_{a,b}$ is Kronecker delta, intra- and inter-plaquette interactions correspond to $a=0$ and $a=1$, and $\langle i,j\rangle$ and $\langle\langle i,j \rangle\rangle$ denote the nearest-neighbor (NN) and the next-nearest-neighbor (NNN) pairs of sites, respectively, as illustrated in FIG. \ref{Fig:Hamiltonian}. Here $J_1$ and $J_2$ are intra-plaquette whilst $\lambda J_1$ and $\lambda J_2$ are inter-plaquette NN and NNN couplings with $\lambda$ being the inter-plaquette coupling strength parameter. 

The model (\ref{eq:Hamiltonian}) with plaquette structure, and in general spin-$1/2$ antiferromagnetic Heisenberg model with various interactions, has been a topic of great interest because of its close connection to cuprate superconductors\cite{Lee,Manousakis,Harland,Altman} and owing to possibility of realization of such models in experiments with cold atoms in optical lattices.\cite{Goral,Lahaye,Nascimbene,Rey,Olschlager} The model was also proposed as a possible model for (CuCl)LaNb$_2$O$_7$.\cite{Kogeyama,Ueda}Additionally models with quadrumerized or various other checkerboard patterns were suggested as models for Bi$_2$Sr$_2$CaCu$_2$O$_{8+\delta}$ and Ca$_{2-x}$Na$_x$CuO$_2$Cl$_2$.\cite{Stock,Tranquada,Shen} 

In these models a quantum phase transition (QPT) from paramagnetic to antiferromagnetic state occurs by competition between antiferromagnetic bonds of different strength as in well studied dimerized Heisenberg models.\cite{Schmidt,Singh, Wenzel2,Jiang,Ma,Fritz,Leite,Merchant} As expected from quantum-to-classical mapping two-dimensional quantum coupled-dimer and coupled-plaquette magnets belong in the same universality class as the class of the three-dimensional classical Heisenberg model referred to as $O(3)$ universality class.\cite{Sachdev3} However in some cases when the dimer pattern lacks a certain symmetry, for example for staggered dimer model also called $J$-$J'$ model which corresponds to Hamiltonian (\ref{eq:Hamiltonian}) in the limit $J_2\gg J_1$ that lacks discrete lattice rotational symmetry, an additional cubic interaction of critical fluctuations not present in the classical $O(3)$ model appears in the low-energy quantum field theory.\cite{Ma,Fritz} This cubic interaction is an interesting quantum effect with no classical counterpart that causes non-monotonic finite-size scaling behavior. Similar non-monotonic scaling also appears for deconfined quantum phase transitions that are characterized by exotic fractionalized quasi-particles and emergent gauge fields.\cite{Sachdev,Sachdev2,Sentil1,Sentil2,Sentil3,Balents,Levin,Sandvik1,Sandvik2,Shao,Ma2,Zhao,You} Moreover $J$-$J'$ model is also connected with Shastry-Sutherland model that explains critical properties of SrCu$_2$(BO$_3$)$_2$.\cite{Shastry,Lauchli,Zayed,Corboz,Zhao2}

In our numerical calculations we use NetKet\cite{Carleo3} and TesorFlow\cite{Abadi} libraries and consider the cases when $J_2=0$ and $J_2=J_1$ for a range of the inter-plaquette coupling strength parameter $\lambda$ values. To identify ordered and disordered phases and quantum phase transitions we write the ground-state wave-function in the form of a RBM and optimize the RBM parameters using RL (VMC) method. For both cases we further calculate the ground-state spin-spin correlation matrices   $M_{ij}=\langle \vec{S}_i\vec{S}_j\rangle$ for a range of inter-plaquette coupling strength parameter $\lambda$ values away from the QPT point and use images of these matrices as input data for a CNN based SML algorithm that detects QPT.

The unfrustrated case with the NNN couplings $J_2=0$ has so far been studied by several methods, exact diagonalization (ED),\cite{Voigt} Ising series expansion,\cite{Singh2} contractor renormalization,\cite{Albuquerque} real space RG,\cite{Fledderjohann} coupled cluster,\cite{Gotze} and quantum Monte Carlo (QMC)\cite{Albuquerque,Wenzel,Xu,Ran} methods. The results for the frustrated case have been calculated by combination of the Lanczos ED and coupled cluster method (CCM) which are both powerful general many-body methods. \cite{Gotze,Kruger} Providing accuracy of the finite-size scaling procedure QMC calculations are expected to give the most accurate results in unfrustrated regime. However in the frustrated regime QMC calculations can not be used to study properties of the system since the method in that case suffers from the minus sign problem. Namely, when partition function of a D dimensional quantum system is expanded in terms of D+1 dimensional classical configurations, weights of these configurations can be both positive and negative or complex and thus invalidate their usual Monte Carlo interpretation as probability distribution.\cite{Hirsch, Loh} Contrary to QMC method SML combined with RBM VMC is a sign problem free method that can provide accurate description of the system phases and phase transitions even in the presence of frustration.

After confirming validity of the RBM ansatz to study ground-state properties of the system for a smaller system size of $N=16$ lattice sites and with periodic boundary conditions for which ED calculations can be done for comparison, we further calculate the ground-state spin-spin correlation matrices for the system sizes of $N=L\times L$ lattice sites with $L=6$, $8$ and $10$. For each system size we calculate 4000 matrices for the values of the inter-plaquette coupling strength parameter $\lambda$ deep in pRVB and AFM phases away from the QPT point. Images are further resized to $L\times L$ pixel values and used as a training (90\% of the images) and validation (10\% of the images) datasets in a CNN based SML algorithm. Trained CNN is then used to classify images of the ground state spin-spin correlation matrices calculated for the remaining interval of $\lambda$ values that contains QPT point $\lambda_c$. SML calculations are performed for hundred random choices of 10\% of the validation data and the final probabilities for the images to belong to pRVB or AFM class calculated as an average over hundred CNN outputs. The point of QPT is identified as the point where pRVB and AFM probabilities are both $50\%$, that is as the point where the network is maximally confused. 

Although consideration of much larger system sizes would be necessary to accurately extrapolate results for finite system sizes to thermodynamic limit ($L\rightarrow \infty$) our results show that the values for $\lambda_c(L)$ rapidly approach $\lambda_c(L\rightarrow \infty)$ values calculated previously by several other methods. We find the best agreement with the results obtained with real space RG method and CCM. For the unfrustrated case with NNN couplings $J_2=0$ previously obtained real space RG and CCM results are $\lambda_c^{RG}\approx 0.4822$\cite{Fledderjohann} and $\lambda_c^{CCM}\approx 0.47$,\cite{Gotze} whilst the QMC and finite-size scaling method result is $\lambda_c^{QMC}\approx 0.5485$.\cite{Albuquerque,Wenzel,Ran} For the frustrated $J_2=J_1$ case the CCM result is $\lambda_c=0.301$.\cite{Gotze} For the largest system size of $N=10\times10$ lattice sites we find the values $\lambda_c\approx 0.465$ and $\lambda_c\approx 0.355$. Whilst the QMC results suggest that the real space RG results converge non-monotonously towards the value obtained with QMC calculations when further excited states with more that one quintuplet are included in the RG calculations,\cite{Fledderjohann} our results indicate that the RG results might converge monotonously towards the value close to the value obtained with the CCM.

The paper is organized as follows. In Sec. \ref{sec:RBM} we describe the RBM ansatz for the ground-state wave-function. In Sec. \ref{sec:VMC} we explain RL scheme for RBM parameters optimization and calculation of the ground-state spin-spin correlation matrices that serve as a training data in the SML algorithm for QPT detection described in Sec. \ref{sec:SML}. We summarize our results, draw conclusions and discuss future research directions in the last section, Sec. \ref{sec:Conculsions}.

\section{Restricted Boltzmann machine representation of the ground-state wave function}
\label{sec:RBM}
\begin{figure}[b!]
\includegraphics[width=\columnwidth]{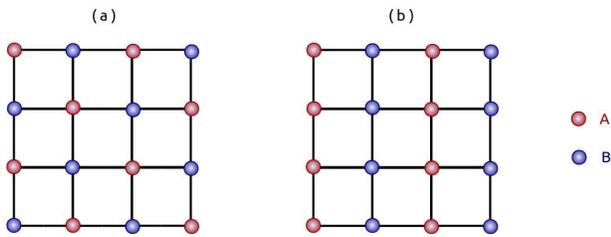}
\caption{\label{Fig:AB_sublattices} Square lattice sublattices (A and B, denoted with red and blue sites) with (a) checkerboard and (b) collinear patterns.
}
\end{figure}

To calculate the ground-state spin-spin correlation matrices used to train SML algorithm described in Sec. \ref{sec:SML} we represent the ground-state wave-function of the system with a restricted Boltzmann machine (RBM) ansatz:
\begin{equation}\label{eq:RBM1}
|\psi_{RBM}\rangle = \sum_{\{\vec{s}\}} (-1)^{n_{A\uparrow}}\psi_{RBM}(\{\vec{s}\})|\{\vec{s}\}\rangle,
\end{equation}
where 
\begin{equation}\label{eq:RBM2}
\psi_{RBM}(\{\vec{s}\})=e^{\sum_{j=1}^Na_js_j}\prod_{i=1}^M\cosh(b_i+\sum_jw_{ij}s_j)
\end{equation}
and $|\{\vec{s}\}\rangle=|s_1,s_2,...,s_N\rangle$ are spin- $1/2$ configurations for $N$ lattice sites. Here we have included basis rotation corresponding to the Marshall's sign rule ( sign $(-1)^{n_{A\uparrow}}$ where $n_{A\uparrow}$ is the total number of $\uparrow$ spins on A subset of the lattice sites) which speeds up the convergence of the VMC calculations by providing a good starting sign structure particularly at low frustration.\cite{Marshall,Auerbach,Schollwock,Choo} For a bipartite lattice typical choice of the sites A corresponds to one of the two sublattices A and B in the bipartite lattice. For a square lattice A and B sublattices usually form checkerboard or collinear patterns as shown in FIG. \ref{Fig:AB_sublattices}. For the two cases that we have studied, $J_2=0$ and $J_2=J_1$ subset A corresponds to sublattice A in FIG. \ref{Fig:AB_sublattices} (a) and FIG. \ref{Fig:AB_sublattices} (b), respectively. 

\begin{figure}[t!]
\includegraphics[width=\columnwidth]{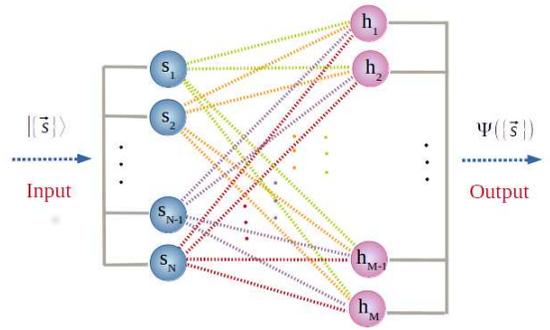}
\caption{\label{Fig:RBM} RBM architecture that is used as an ansatz for the ground-state wave-function. The network consists of one visible layer with $N$ visible artificial neurons ($s_1$,...,$s_N$) and one hidden layer with $M$ hidden artificial neurons ($h_1$,...,$h_M$) with interactions only between the visible and hidden units.  Network inputs are visible spin configurations $|\{\vec{s}\}\rangle\equiv|s_1,...,s_N\rangle$ and as an output the network calculates probability of a visible configuration $P(\{\vec{s}\})$ equivalent to the restricted Boltzmann machine ansatz coefficients $\psi_{RBM}(\{\vec{s}\})=P(\{\vec{s}\})$.
}
\end{figure}
RBM ansatz that encodes a many-body quantum state of spin-$1/2$ quantum system with $N$ spins is constructed of one visible layer with $N$ nodes (neurons, visible spins) that correspond to physical qubits ($s_1$, $s_2$, ... ,$s_N$), and one hidden layer with $M$ auxiliary spin variables ($h_1$, $h_2$, ... , $h_M$) as shown in FIG. \ref{Fig:RBM}. Two sets of binary units (classical spins) that correspond to visible ($s_i$, $i=1,...,N$) and hidden ($h_i$, $i=1,...,M$) neurons in Boltzmann machine architectures interact through an Ising interaction and corresponding Hamiltonian H is defined as 
\begin{eqnarray}\label{eq:H_RBM}
H=\sum_{j=1}^N a_js_j+\sum_{i=1}^M b_ih_i + \sum_{i<j} c_{ij}s_is_j \\
+ \sum_{i,j}w_{ij}h_iv_j+\sum_{i<j}d_{ij}h_ih_j. \nonumber
\end{eqnarray}
Joint probability distribution over visible and hidden units is then defined as the Boltzmann weight of this Hamiltonian:
\begin{equation}\label{eq:Boltzmann_weight}
P(\{\vec{s}\},\{\vec{h}\})=\frac{1}{Z}e^{H(\{\vec{s}\},\{\vec{h}\})}
\end{equation}
where $Z$ is the partition function $Z=\sum_{\{\vec{s}\},\{\vec{h}\}}e^{H(\{\vec{s}\},\{\vec{h}\})}$. Then marginal probability of a visible configuration can be obtained by summing over all possible hidden configurations
\begin{equation}\label{eq:P_visible}
P(\{\vec{s}\})=\sum_{\{\vec{h}\}}\frac{1}{Z}e^{H(\{\vec{s}\},\{\vec{h}\})}.
\end{equation}
Here we take visible spin configurations $|s_1,...,s_N\rangle$ as network inputs and interpret the wave-function as complex probability distribution that the network tries to approximate. Therefore the Boltzmann machine ansatz wave-function coefficients are 
\begin{equation}\label{eq:wf_coefficients}
\psi_{BM}(\{\vec{s}\})=P(\{\vec{s}\}).
\end{equation}

For the RBM ansatz Hamiltonian (\ref{eq:H_RBM}) includes only interactions $w_{ij}$ between the visible and hidden units ($c_{ij}=0$, $d_{ij}=0$) and biases $a_j$ and $b_j$. Sum over the hidden spins in (\ref{eq:P_visible}) can then be performed analytically and the wave-function ansatz coefficients written as
\begin{equation}\label{eq:coefficients_RBM}
\psi_{RBM}(\{\vec{s}\})=e^{\sum_{j=1}^Na_js_j}\prod_{i=1}^M\cosh(b_i+\sum_jw_{ij}s_j).
\end{equation}

ANN coefficients are in general taken to be complex in order to describe both the amplitude and the phase of the wave-function. Representability theorems\cite{Kolgomorov,Hornik1,Hornik2,LeRoux} guarantee existence of an ANN approximation for every sufficiently smooth and regular high-dimensional function, allowing ANN description of complicated many-body wave-functions. Quality of the ANN approximation can be systematically improved by increasing hidden variable density $\rho=M/N$ where $M$ is the number of hidden and $N$ the number of visible neurons (spins), respectively. 

In one dimension, for well studied matrix product state (MPS) ansatz hidden variable density $\rho$ is analogous to the bond dimension, that is dimension of the matrices in the MPS ansatz, which determines the space complexity of the MPS. However, unlike MPS ansatz, ANN ansatz is well suited for description of quantum states in arbitrary dimensions due to intrinsically nonlocal correlations in space generated by hidden units. In our calculations we take $\rho=4$ which shows to be sufficient to obtain quite accurate estimates of the ground state properties of the system. 

The parameters of the ANN are obtained using VMC method equivalent to a reinforcement learning scheme in which the ground-state wave-function is learned on the basis of feedback from variational principle as described in the following section.

\section{Variational Monte Carlo optimization of the restricted Boltzmann machine parameters - reinforcement learning of the ground-state wave-function based on feedback from the variational principle}
\label{sec:VMC}

Parameters of the RBM ansatz can be optimized (trained) using VMC method through minimization of the energy expectation value 
\begin{equation}\label{eq:Energy}
E=\frac{\langle\psi_{RBM}|H|\psi_{RBM}\rangle}{\langle\psi_{RBM}|\psi_{RBM}\rangle}
\end{equation}
with respect to the network parameters ($a_j$,$b_j$,$w_{ij}$), $i,j=1,...,N$, with N being the number of lattice sites.

Energy $E$ can be optimized using several optimization approaches. Here we use stochastic reconfiguration (SR) method introduced by Sorella \emph{et al.}\cite{Sorella2,Becca} which can be viewed as an effective approximate imaginary time evolution in the variational subspace of the full Hilbert space. The method was initially developed to stabilize and partially solve the sign instability in the Green function Monte Carlo technique for continuous models\cite{Sorella} and then as an efficient and robust optimization scheme for lattice systems\cite{Sorella1}, small atoms\cite{Casula1} and molecules\cite{Casula2} that allows energy expectation value minimization for a many-parameter variational ansatz wave-function in an arbitrary functional form. SR method uses more information about the variational ansatz wave-function than other common simpler optimization protocols such as stochastic gradient descent (SGD) and therefore allows faster optimization of the many-body ansatz wave-function and more reliable convergence. Instead of optimizing energy directly the method at each step tries to maximize the overlap between the ansatz and the result of its imaginary time evolution. 
\begin{figure}[t!]
\includegraphics[width=\columnwidth]{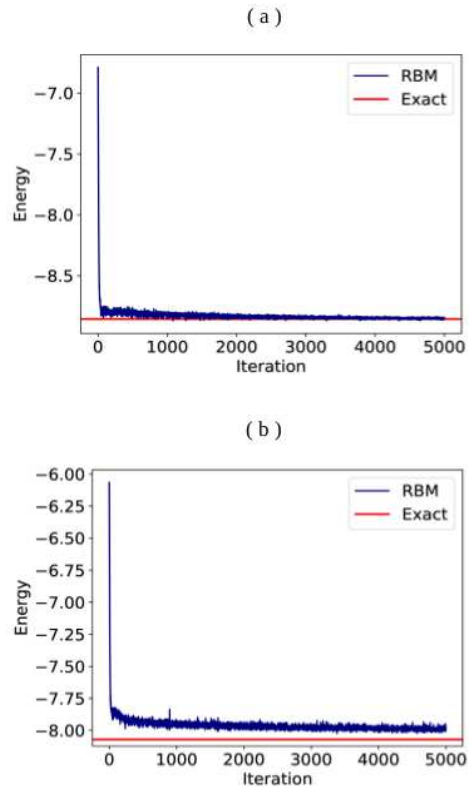}
\caption{\label{Fig:E0_RBM} Monte Carlo energy estimate obtained with variational optimization of the RBM ansatz with hidden unit density $\rho=4$ for the system size of $N=4\times 4 =16$ lattice sites and with periodic boundary conditions. Here the inter-plaquette coupling strength parameter $\lambda=0.5$ and figures (a) and (b) correspond to the cases $J_2/J_1=0$ and 1, respectively. Energy is calculated at each iteration step. Monte Carlo energy estimate converges to the exact ground state energy (red lines) up to a relative error $100\times(E^{RBM}-E^{Exact})/E^{Exact}$. 
}
\end{figure}

For an ansatz wave-function $|\psi(\alpha_1,\alpha_2,...,\alpha_\mathcal{N})\rangle$ with a set of variational parameters $\{\alpha_1,\alpha_2,...,\alpha_\mathcal{N}\}$ the variational subspace is spanned by the wave-function and its derivatives with respect to each variational parameter $|\psi^i\rangle \equiv \frac{\partial |\psi\rangle}{\partial \alpha_i}$, $i=1,...,\mathcal{N}$. Within the SR method the energy of the ansatz wave-function is minimized by repeatedly applying imaginary time evolution operator $T=e^{-\tau H}\approx 1-\tau H$ to the wave-function where $\tau$, that corresponds to the learning rate in ML language, is taken to be a small positive number and operator $T$ expanded to the first order in $\tau$. After each application of  $T$  the result is projected onto variational subspace to obtain a new function of the same form and a new set of variational parameters calculated using Monte Carlo sampling. Namely, after application of the evolution operator $T\approx 1-\tau H$ variational parameters $\{\alpha_1,\alpha_2,...,\alpha_\mathcal{N}\}$ are replaced by a new set of parameters $\{\alpha'_1,\alpha'_2,...,\alpha'_\mathcal{N}\}$ such that $|\psi(\alpha'_1,...,\alpha'_\mathcal{N})\rangle \equiv |\psi'\rangle $ is a good approximation to $e^{-\tau H}|\psi(\alpha_1,...,\alpha_\mathcal{N})\rangle \approx (1-\tau H) |\psi(\alpha_1,...,\alpha_\mathcal{N})\rangle \equiv |\psi_\tau^{'}\rangle$, that is, projection of $|\psi'_\tau\rangle$ to variational subspace. 
\begin{figure}[b!]
\includegraphics[width=\columnwidth]{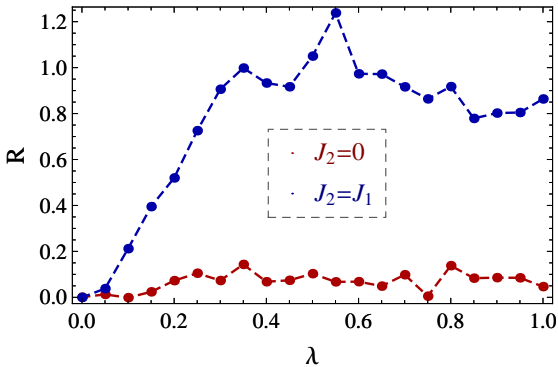}
\caption{\label{Fig:Rerr} Relative error $R=100\times(E_0^{RBM}-E_0^{ED})/E_0^{ED}$ for the ground-state energy of the Hamiltonian (\ref{eq:Hamiltonian}) obtained by RBM optimization with respect to the exact ground state energy (ED result) for $J_2=0$ and $J_2=J_1$ (blue and red data points, respectively) and a range of values of the inter-plaquette coupling parameter $\lambda$. Here results are for the system size of $N=4\times4=16$ lattice sites with periodic boundary conditions. RBM hidden unit density is $\rho=4$.
}
\end{figure}

\begin{figure}[t!]
\includegraphics[width=\columnwidth]{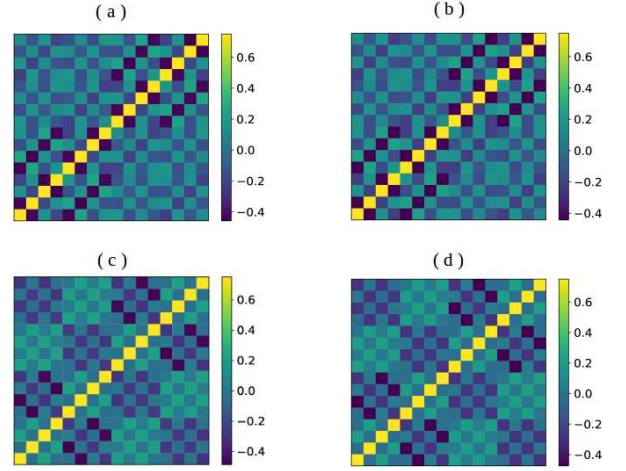}
\caption{\label{Fig:sscf} The ground-state spin-spin correlation matrices $M_{ij}=\langle \vec{S}_i\vec{S}_j\rangle$ for the inter-plaquette coupling strength parameter $\lambda=0.5$ in Hamiltonian (\ref{eq:Hamiltonian}) and for the system size  of $N=4\times4$ lattice sites with periodic boundary conditions. Figures (a) and (b) correspond to the unfrustrated case $J_2=0$, whilst (c) and (d) correspond to the frustrated case with $J_2=J_1$. Matrices in figures (a) and (c) are obtained using RBM based VMC method and matrices in figures (b) and (d) using ED method. RBM hidden artificial neuron density is $\rho=4$. 
}
\end{figure}

$|\psi'\rangle$ can be found by maximizing the overlap of the unnormalized wave-functions $|\psi'\rangle$ and $|\psi'_{\tau}\rangle$:
\begin{equation}\label{eq:overlap}
|D|^2=\frac{\langle\psi'_{\tau}|\psi'\rangle\langle\psi'|\psi'_{\tau}\rangle}{\langle\psi'_{\tau}|\psi'_{\tau}\rangle\langle\psi'|\psi'\rangle},
\end{equation}
that is, from the conditions:
\begin{equation}\label{eq:overlap_max}
\frac{\partial |D|^2}{\partial \alpha_i}=0 \; \; \; \forall i. 
\end{equation}
Writing new variational parameters $\{\alpha'_1,\alpha'_2,...,\alpha'_\mathcal{N}\}$ as $\alpha'_{i}=\alpha_i+\delta\alpha_{i}$ ($i=1,...,\mathcal{N}$) and keeping only terms linear in $\tau$ and $\delta\alpha_i$ ($\forall i$) the conditions (\ref{eq:overlap_max}) can be rewritten as:
\begin{eqnarray}\label{eq:overlap_max2}
&&\sum_j \delta \alpha_j \left[\frac{\langle\psi^j|\psi^i\rangle}{\langle\psi|\psi\rangle}-\frac{\langle\psi^j|\psi\rangle}{\langle\psi|\psi\rangle}\frac{\langle\psi|\psi^i\rangle}{\langle\psi|\psi\rangle}\right]= \\
&& \tau\left[\frac{\langle\psi|H|\psi^i\rangle}{\langle\psi|\psi\rangle}-\frac{\langle\psi|H|\psi\rangle}{\langle\psi|\psi\rangle}\frac{\langle\psi|\psi^i\rangle}{\langle\psi|\psi\rangle}\right]\nonumber,
\end{eqnarray}
where $|\psi\rangle\equiv |\psi(\alpha_1,\alpha_2,...,\alpha_\mathcal{N})\rangle$ and $|\psi^i\rangle=\partial|\psi\rangle/\partial\alpha_i$. 

To efficiently evaluate $\left\{\delta\alpha_1,...,\delta\alpha_{\mathcal{N}}\right\}$ numerically energy and all other expectation values in Eq. (\ref{eq:overlap_max2}) are written as Monte Carlo averages with respect to quantum probability distribution 
\begin{equation}\label{eq:probability_distribution}
P(\left\{\vec{s}\right\})= \frac{|\langle\left\{\vec{s}\right\}|\psi\rangle|^2}{\langle\psi|\psi\rangle}
\end{equation}
by inserting resolution of identity summation of all possible spin configurations $\sum_{\{\vec{s}\}} |\{\vec{s}\}\rangle\langle\{\vec{s}\}|=\mathbbm{1}$ in the expectation values in Eq. (\ref{eq:overlap_max2}) where $|\{\vec{s}\}\rangle=|s_1,...,s_{\mathcal{N}}\rangle$ are $s_i^z$ basis states. The expectation values are then written as 
\begin{equation}\label{eq:evH}
\frac{\langle\psi|H|\psi\rangle}{\langle\psi|\psi\rangle}=\sum_{\{\vec{s}\}}P(\{\vec{s}\})E(\{\vec{s}\})
\end{equation}
and
\begin{equation}\label{evDH}
\frac{\langle\psi|H|\psi^i\rangle}{\langle\psi|\psi\rangle}=\sum_{\{\vec{s}\}}P(\{\vec{s}\})E^{*}(\{\vec{s}\})O_i(\{\vec{s}\}),
\end{equation}
where 
\begin{equation}\label{eq:Eloc}
E(\{\vec{s}\})=\frac{\langle\{\vec{s}\}|H|\psi\rangle}{\langle\{\vec{s}\}|\psi\rangle}
\end{equation}
is the local energy and 
\begin{equation}
O_i(\{\vec{s}\})=\frac{\langle \{\vec{s}\}|\psi^i\rangle}{\langle\{\vec{s}\}|\psi\rangle}=\frac{\partial}{\partial \alpha_i}\ln \langle \{\vec{s}\}|\psi\rangle.
\end{equation}
The Eq. (\ref{eq:overlap_max2}) further becomes
\begin{equation}\label{eq:max_overlap3}
\sum_{j}S_{ij}\delta\alpha_j=-\tau F_i
\end{equation}
with the covariance matrix $S$ and generalized forces $F_i$ defined as 
\begin{equation}\label{eq:Sij}
S_{ij}=\langle O_j^{*}O_i\rangle-\langle O_j^{*}\rangle\langle O_i\rangle
\end{equation}
and 
\begin{equation}\label{eq:Fi}
F_i=\langle E_{loc}^{*}O_i\rangle-\langle E_{loc}^{*}\rangle\langle O_i\rangle.
\end{equation}
SR updates for coefficients $\alpha\equiv\{\alpha_1,\alpha_2,...,\alpha_\mathcal{N}\}$ at the $n$-th iteration are therefore of the form
\begin{equation}\label{eq:SR_update}
\alpha(n+1)=\alpha(n)-\tau S^{-1}(n)F(n).
\end{equation}
In the case of non-invertible covariance matrix $S^{-1}$ is replaced with Moore-Penrose pseudo-inverse or a positive constant is added to the diagonal $S$ matrix elements to obtain regularized matrix that is invertible. 

To check validity of the RBM ansatz for the ground-state wave-function we first calculate the ground-state energy and spin-spin correlation matrices $M_{ij}=\langle\vec{S}_i\vec{S}_j\rangle$ for a small system size of $N=4\times4=16$ lattice sites with periodic boundary conditions and compare with the exact results obtained using ED method. ED and RBM optimization results (with $\tau=0.05$ in SR) are shown in FIG. \ref{Fig:E0_RBM} - FIG. \ref{Fig:sscf}. In our calculations relatively small hidden unit density $\rho=4$ for the RBM ansatz already provides the same accuracy of the results as obtained for other best known ansatz wave-functions, for example projected entangled-pair states (PEPS) or EPS.\cite{Carleo,Wang2,Mezzacapo,Mezzacapo2}

Monte Carlo energy estimate converges to the exact ground-state energy up to a relative error that is even in the frustrated regime smaller than $1.3\%$ for a wide range of values of the inter-plaquette coupling strength parameter $\lambda$. For the non-frustrated case the relative error is smaller than $0.2\%$.

We also note that the optimized ansatz for the frustrated case $J_2=J_1$ does not have discrete lattice reflection symmetry in the line $x=y$ defined as R: $(x,y)\rightarrow (y,x)$ whilst the ground-state eigenstate of the Hamiltonian obtained by ED method does have the reflection  symmetry. Therefore, to compare ED and RBM results for $J_2=J_1$, we calculate the ground state spin-spin correlation matrix in FIG. \ref{Fig:sscf} (c) with the symmetrized RBM ansatz wave-function $|\psi^R_{RBM}\rangle=|\psi_{RBM}\rangle+\mathbb{R}|\psi_{RBM}\rangle$. The matrix elements are then $M_{ij}^R=\langle\psi_{RBM}|\vec{S}_i\vec{S}_j|\psi_{RBM}\rangle + \langle\psi_{RBM}|\vec{S}_{R(i)}\vec{S}_{R(j)}|\psi_{RBM}\rangle $.

\section{Supervised machine learning of the antiferromagnet to plaquette resonating valence bond state quantum phase transitions}
\label{sec:SML}

\begin{figure}[b!]
\includegraphics[width=\columnwidth]{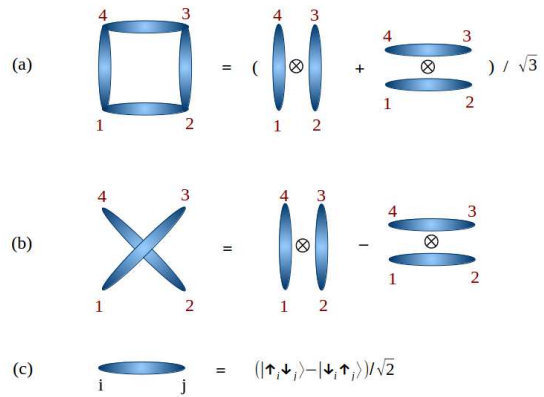}
\caption{\label{Fig:pRVB} A singlet bond (c) and pRVB states with (a) s-wave and (b) d-wave symmetry. The s-wave pRVB state is symmetric with respect to an exchange of two diagonal spins, whilst d-wave pRVB state is antisymmetric.  
}
\end{figure}

\begin{figure}[t!]
\includegraphics[width=\columnwidth]{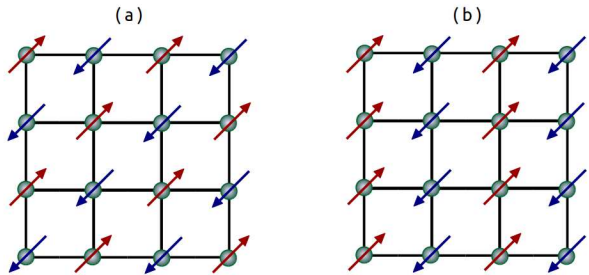}
\caption{\label{Fig:AFM} Illustration of the classical ($s\rightarrow \infty$) phases for (a) $J_2=0$ with N\'{e}el antiferromagnetic order and (b) $J_2=J_1$ with collinear striped antiferromagnetic order.  
}
\end{figure}

We further use SML approach to distinguish phases and locate QPTs for both frustrated and unfrustrated parameter regimes, in particular for $J_2=J_1$ and $J_2=0$. Here a phase transition occurs at some critical value $\lambda_c$ of the inter-plaquette coupling strength parameter $\lambda$. For $\lambda < \lambda_c$ the system is in a paramagnetic plaquette resonating valence bond (pRVB) phase and for $\lambda > \lambda_c$ the system exhibits antiferromagnetic ordering. When NNN coupling $J_2=0$ pRVB phase has s-wave symmetry and corresponding AFM phase N\'eel order (NAFM) whilst for the frustrated case with $J_2=J_1$ pRVB phase has d-wave symmetry and is separated by a phase transition from corresponding AFM phase with collinear striped long range order (CAFM). Single pRVB plaquette and schematic representation of classical AFM phase for both cases are illustrated in FIG. \ref{Fig:pRVB} and FIG. \ref{Fig:AFM}.

In our SML calculations gray-scale images of spin-spin correlation matrices $M_{ij}=\langle\vec{S}_i\vec{S}_j\rangle$ in AFM and pRVB part of the phase diagram are placed in two classes labeled by 0 and 1, respectively, and used as input data to train a CNN architecture that after training procedure predicts labels (classes) of new images not included in the training data set and identifies position of the QPT. Training procedure corresponds to the CNN parameters optimization with respect to a loss function here taken to be the binary cross-entropy loss function: 
\begin{equation}\label{eq:binary_cross_entropy}
L = - \frac{1}{N_t}\sum_{i=1}^{N_t}[y_i\log P(y_i) + (1 - y_i)\log(1-P(y_i)]
\end{equation}
where $\log$ is a base $2$ logarithm, $y_i$ is assigned label ($1$ or $0$ for two classes), $P(y_i)$ is the predicted probability of the input from training data set labeled by $y_i$ being in class 1 and $N_t$ is the number of training data examples. The loss function decreases as the predicted probability approaches the value of the actual label and gives information how good the CNN is to make predictions on new images not included in the training data set.

Cross-entropy is a measure from information theory that calculates difference between two probability distributions for a given random variable or set of events by calculating total entropy between two distributions. Low probability surprising events contain more information whilst high probability unsurprising events contain less information where information is defined as $-\log(P(x))$ with $P(x)$ being the probability of an even $x$. For a random variable $x$ with $N_t$ discrete states (data set of size $N_t$) the entropy, which is defined as the number of bits required to encode and transmit an event randomly selected from a probability distribution, is then $-\sum_{x\in N_t}P(x)\log(P(x))$. 
\begin{figure*}[t!]
\includegraphics[width=\textwidth]{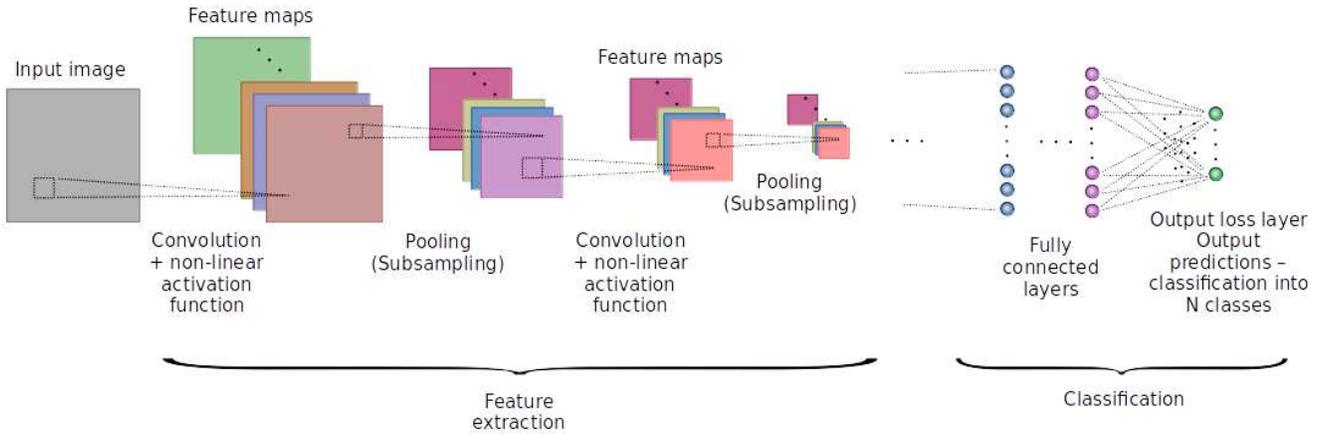}
\caption{\label{Fig:CNN} A typical CNN architecture consists of convolutional layers interspersed with pooling layers and then followed by fully connected layers and an output classification layer. CNN input is an image that the network attempts to classify, convolutional layers extract features from the input image, pooling layers coarse-grain spatial information by performing sub-sampling, fully connected layers combine features obtained from previous convolutional and pooling layers and evaluate classification decision and the output layer predicts a class of the input image.  
}
\end{figure*}

On the other hand cross-entropy calculates average number of bits needed to encode data from original distribution $P$ compared to another model distribution $Q$ that approximates distribution $P$. The cross-entropy then represents the number of additional bits to represent an event using $Q$ instead of $P$ and can be calculated as $-\sum_{x\in N}P(x)\log(Q(x))$. For classification tasks class labels are known and corresponding probability distribution has no information since the outcome is certain which means that the entropy of a known class label is always zero. When two distributions are identical the cross-entropy between them is equal to the entropy for the probability distribution, which means that the cross-entropy of real and predicted probability distributions for a class label is also zero. When cross-entropy loss function is calculated for a training data set the cross-entropy is averaged over all examples in the dataset and zero loss therefore indicates that 
the predicted class probabilities are identical to the probabilities in the training dataset. Binary cross-entropy can thus be used as a loss function for a binary classification task with only two classes. 

During the training process the loss function (\ref{eq:binary_cross_entropy}) is optimized using mini-batch SGD method. Within gradient descent (GD) optimization method values of the CNN parameters that minimize a loss function are found iteratively by updating each parameter (weight) $\alpha$ in the loss function in each iteration step $n$ according to the equation:
\begin{equation}\label{eq:GD}
\alpha^{n+1}=\alpha^n -\nu \frac{\partial L}{\partial \alpha},
\end{equation}
where $\nu$ is a step size hyper-parameter called learning rate in ML. The learning rate is one of the most important parameters that requires careful tuning when training an ANN to achieve fast convergence. This hyper-parameter controls the speed at which the ANN model learns. If the learning rate is too small optimization will be very slow and many iterations will be necessary to reach the minimum of the loss function, and if the rate is too large the training may not converge at all or can even diverge since weight changes can be so large that the optimizer overpasses the minimum causing an increase of the loss function value. 

The mini-bath SGD algorithm is computationally more efficient than GD algorithm and allows more robust convergence. In GD optimization all training data set samples are passed through the network for a single update of the network weights. In mini-batch SGD learning algorithm\cite{Sra} instead of the whole data set a randomly selected subset of data is needed for each update where the true gradient is approximated by stochastic approximation of the gradient calculated from that randomly selected subset of data. The size of the subset is called batch size. One complete pass of all training samples through the network is called an epoch and training step usually requires more than a few epochs. 

One of the most common problems in ML training step is over-fitting. Over-fit models fit very well training data. However such models are not able to classify well new data not included in the training data set. In other words over-fitting leads to a model that is very well trained however unable to generalize, that is unable to correctly predict the label of a new input sample. The model just memorizes the training data and the noise and has high variance and low bias. Variance reflects model dependence on training data set and bias reflects assumptions made about the training data. Contrary to over-fit models an under-fit model has low variance and high bias. In that case the model makes strong assumptions about data and fails to capture underlying pattern in data. Over-fitting and under-fitting can be identified by calculating validation and training accuracies. 

A validation data set is a set of data for which the labels that an ANN tries to predict are known and which is excluded from the training data set. Depending on the size of the available data set for which the labels are known a validation set is   formed from 10-30$\%$ of the data while the rest of the data is used for training. The accuracy of a ML classification algorithm is defined as 
\begin{equation}\label{eq:accuracy}
A=\frac{N_c}{N_{tot}},
\end{equation}
where $N_c$ is the number of correct predictions and $N_{tot}$ is the total number of predictions. If the accuracies for the training and validation data sets are $A_t$ and $A_v$ over-fitting happens when $A_t > A_v$ and under-fitting when 
$A_t < A_v$. Optimally $A_t\approx A_v\approx 1$. Whilst increasing the model complexity can help with under-fitting there are several methods to reduce over-fitting, like increasing the data set size, decreasing complexity of the model, or applying various regularization techniques.\cite{Sra} In our SML calculations we set the batch size to 32, and find that the loss function (\ref{eq:binary_cross_entropy}) approaches zero and $A_t$, $A_v\approx 1$  already after ten training epochs.

A typical CNN architecture consists of three kinds of layers: convolutional layers, pooling layers and fully connected layers as illustrated in FIG. \ref{Fig:CNN}. 

An input gray-scale image can be represented as a matrix of pixel values where the value of each pixel in the matrix ranges from 0 to 255 with zero indicating black and 255 indicating white. A color image can be represented with three 2D matrices each having pixel values in range 0 to 255, one for each color - red, green or blue. A color image is said to have three channels (each channel representing certain component of an image), whilst a gray-scale image has just one channel. 

In image processing a kernel or convolution matrix is a small matrix used for achieving a wide range of effects on original image, for example blurring, sharpening, embossing, and edge detection. In CNN architectures and ML such kernels are used for feature extraction, that is for determining the most important parts of an image, the process more generally referred to as convolution. Convolution is the operation of adding each element of the image to its local neighbor weighted by the kernel. The general form for convolution (*) of two matrices A and B is:
\begin{eqnarray}\label{eq:MatrixConvolution}
&&\begin{bmatrix}
A_{11} & A_{12} & . & . & . &A_{1n} \\
A_{21} & A_{22} &   &   &   &A_{2n}\\
     . & .      & . &   &   & . \\
     . & .      &   &.  &   & . \\
     . & .      &   &   & . & .\\
A_{m1} & A_{m2} & . & . & . &A_{mn} 
\end{bmatrix}
*
\begin{bmatrix}
B_{11} & B_{12} & . & . & . &B_{1n} \\
B_{21} & B_{22} &   &   &   &B_{2n}\\
     . & .      & . &   &   & . \\
     . & .      &   &.  &   & . \\
     . & .      &   &   & . & .\\
B_{m1} & B_{m2} & . & . & . &B_{mn} 
\end{bmatrix} \nonumber \\
&=&\sum_{i=0}^{m-1}\sum_{j=0}^{n-1}A_{(m-i)(n-j)}B_{(1+i)(1+j)}
\end{eqnarray}

\begin{figure}[t!]
\includegraphics[width=\columnwidth]{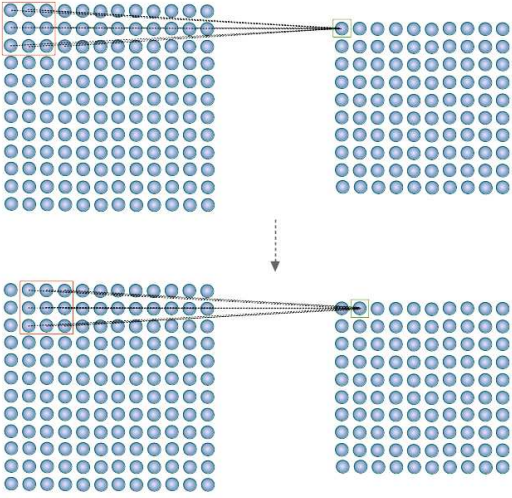}
\caption{\label{Fig:Convolution} Kernel (convolution matrix, feature detector) is a matrix of weights which are multiplied with the input to extract relevant features. A small kernel matrix is slid across the image and convolved with a restricted area of input data (typically of a square shape). Here input image is a gray-scale image with only one channel. The region in the input image connected to a hidden neuron is called local receptive field for the hidden neuron. The process of finding features within a certain window of data is continued until the entire image is covered and a feature map is obtained which is represented by simplified set of pixel values.
}
\end{figure}

Within a convolution step in CNN architecture a small kernel matrix is slid across the image and convolved with small squares of input data as shown in FIG. \ref{Fig:Convolution}. In ML algorithm CNN can learn kernel values that can extract important features. Since image features are learnt using small portions (squares) of input data convolution preserves spatial relationship between pixels.  The region of the input image connected to a hidden neuron is called local receptive field for the hidden neuron. If the local receptive field is moved by S pixels at a time we say a stride length of S is used. In FIG. \ref{Fig:Convolution} stride length is $S=1$. In practice stride lengths of $S\geq 3$ are rarely used. Also, output volume spatial size can be controlled by padding the input with zeros on the border of the input volume. In our calculations kernel matrix is a $3\times 3$ matrix and the stride length is set to $S=1$. 

Whilst for a two-dimensional (2D) convolutions, kernel is a $\mathcal{H}\times \mathcal{W}$ matrix, a filter is defined as a concentration of multiple kernels, each kernel assigned to a particular channel of the input. For $k$ input channels, the filter dimension is therefore $k\times \mathcal{H}\times \mathcal{W}$.  A general convolutional layer in a CNN architecture consists of multiple such filters. Filters' shapes and sizes are chosen based on the dataset to create abstractions and recognize features at the proper scale. 
Each hidden neuron has a bias and weights connected to its local receptive field as illustrated in FIG. \ref{Fig:Convolution}. It is important to note that within a CNN the same weights and bias are used for each hidden neuron in a given feature map, that is, for $j,k$-th hidden neuron the output is:
\begin{equation}
\sigma\left(b+\sum_{l=0}^{\mathcal{W}}\sum_{m=0}^{\mathcal{H}} w_{l,m}a_{j+l,k+m}\right)
\end{equation}
Here $w_{l,m}$ is an array of shared weights, $b$ is shared bias, $a_{x,y}$ is the input activation (value) at position $x,y$ and $\sigma$ is the neural network activation function like, for example, sigmoid function $\sigma(x)=1/(1+e^{-x})$ or rectified linear unit (ReLU) function $\sigma(x)=\max(0,x)$. In practice ReLU activation is often preferred to other activation functions because it allows faster neural network training without significant penalty to generalization accuracy. In our calculations activation function for all neurons in convolutional layers and fully connected layers is ReLU, whereas activation function of the output layer neurons is the softmax function $\sigma(\vec{o})=e^{o_i}/\sum_{i=0}^{N-1} e^{o_i}$ where $\vec{o}=\{o_0,...,o_{N-1}\}$ is the output vector of the network with $N$ output classes and $N=2$ for the problem that we have studied. The softmax function is used as an activation for the output layer to normalize total probability over predicted output classes to one. 

\begin{figure}[t!]
\includegraphics[width=\columnwidth]{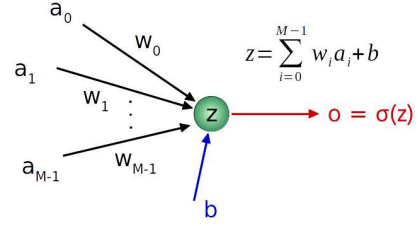}
\caption{\label{Fig:Neuron}Schematic representation of an artificial neuron. Neurons consist of a linear transformation that weights the importance of various inputs and a non-linear activation function $\sigma$ which allows the network consisting of these neurons to capture more complex patterns. Neurons are generally arranged in layers with the output of one layer serving as the input to the next layer.  
}
\end{figure}

In other words, each neuron, illustrated in FIG. \ref{Fig:Neuron}, is a mathematical function that multiplies inputs by weights, adds them together, adds a bias $b$ which can help better fit the data, and then passes the sum to a non-linear (activation) function to become the neuron's output. A non-linear function is applied by a neuron to introduce non-linear properties in the network and allow the network to capture more complex patterns. 
\begin{figure}[b!]
\includegraphics[width=\columnwidth]{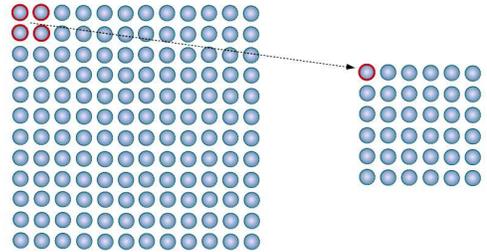}
\caption{\label{Fig:Pooling} Pooling step in a CNN architecture. A pooling layer reduces dimensions of each feature map output from the previous convolutional layer and creates condensed feature maps whilst keeping the most important information. For example max pooling step for $2\times2$ clusters replaces the clusters with the maximum value from each cluster. To avoid information loss we exclude pooling layers from the convolutional neural network in our  calculations.  
 }
 \end{figure}

A parameter sharing scheme used in CNNs is based on assumption that if a feature is useful to compute in a region of input (local receptive field) at some spatial position, then it should also be useful to compute at other positions. Such parameter sharing contributes to the translation invariance of the CNN architectures. 

For image recognition more than one feature map is needed and a complete convolutional layer, which is the core building block of a CNN, consists of several different feature maps where each feature map is defined by a set of shared weights and a shared bias. Depth of a convolutional layer corresponds to the number of filters in that layer. The number of feature maps necessary in a given convolutional layer depends on the number of available examples and task complexity. 

To summarize, the primary purpose of convolution is to extract features from the input image so that CNN can learn the values of feature filters (analyze important features of the image) on its own during the training process. More filters correspond to more image features extracted which allows the network to better recognize patterns in unseen images.
\begin{figure}[t!]
\includegraphics[width=\columnwidth]{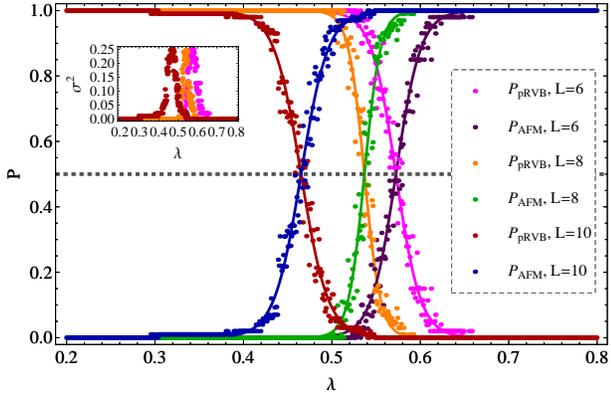}
\caption{\label{Fig:Pswave} Probabilities of the s-wave pRVB ($P_{pRVB}$) and NAFM phases ($P_{AFM}=1-P_{pRVB}$) for the unfrustrated case with the NNN coupling $J_2=0$ in the Hamiltonian (\ref{eq:Hamiltonian}). The probabilities are calculated as averages over hundred CNN output label predictions trained in each case with slightly different training data set. Corresponding variances of the predictions are shown in inset. Standard error for each prediction is $\sigma/\sqrt{n}$ with $n=100$. The results are calculated for the system sizes $N=L\times L$ lattice sites with $L=6,8,10$ and with periodic boundary conditions using SML algorithm with the CNN architecture consisting of three convolutional layers with 256 filters and one fully connected layer with 256 neurons. The output layer has two neurons, one for each phase class. The ground-state spin-spin correlation matrices are calculated with RBM based RL algorithm  in the inter-plaquette coupling strength parameter $\lambda\in\left[0,0.2\right]$ and $\lambda\in\left[0.8,1.0\right]$ intervals (2000 matrices in each interval) that do not contain the transition point and random 10\% of data separated as a validation data set. The remaining 90$\%$ of data is then used to train the network. The network then predicts class labels for 600 matrices calculated in the interval $\lambda\in\left[0.2,0.8\right]$ and detects the QPT which corresponds to the point where the network is maximally confused ($P_{pRVB}=P_{AFM}=0.5$). The solid lines are fits to a logistic function (smooth approximation to the Heaviside step function) $f(x)=1/(1+e^{-k(x-x_0)})$ and $1-f(x)$. For the studied system sizes $\lambda_c (L) \approx 0.573$, $0.537$ and $0.465$  for $L=6$, $8$ and $10$, respectively. 
}
\end{figure}

In addition to convolutional layers CNNs often also contain pooling layers, usually a pooling layer after each convolutional layer. The role of a pooling layer is to simplify the output from the convolutional layer and create condensed feature maps by applying spatial pooling (also called sub-sampling or down-sampling) to each feature map separately. Pooling attempts to reduce the dimensions of each feature map while keeping the most important information. There are several different types of pooling, for example max, average and sum pooling. In a pooling step small regions of output from a convolutional layer are replaced by maximum, average or sum of values in each region as illustrated in FIG. \ref{Fig:Pooling}. Pooling layers therefore coarse-grain convolutional layer output while maintaining spatial structure and locality. This pooling step is very similar to RG decimation step.\cite{Mehta,Janusz,Kadanoff1,Kadanoff2,Efrati,Iso,Lin} 

Translational invariance built in CNN architecture, that is based on assumption that relative positions of various features are more important than exact locations of each feature, is achieved by a combination of convolutional and pooling layers. However a major issue of adding pooling layers in a CNN is that pooling can result in excess information loss. Namely, CNN with pooling layers fails to learn precise spatial correlations between different features and can thus predict inaccurate class label if all features are present even if those features are at wrong spatial positions. Therefore discarding pooling layers in CNN architecture can sometimes lead to better results. To avoid information loss in our SML calculations we thus exclude pooling layers.

Each convolutional and pooling step is a hidden layer and such hidden layers in a CNN architecture are usually followed by fully connected hidden layers and a final output layer. Whilst in a convolutional layer neurons receive input only from a subarea (receptive field) of the previous layer in a fully connected layer each neuron is connected and receives input from every neuron of the previous layer. Although high-level features obtained from convolutional and pooling layers may be sufficient for classifying an input image into different classes based on the training data set, non-linear combinations of such features might be better for the image classification. Purpose of the  fully connected layer is therefore to learn non-linear feature combinations relevant for the classification task, that is prediction of the best label for an input image. 
\begin{figure}[b!]
\includegraphics[width=\columnwidth]{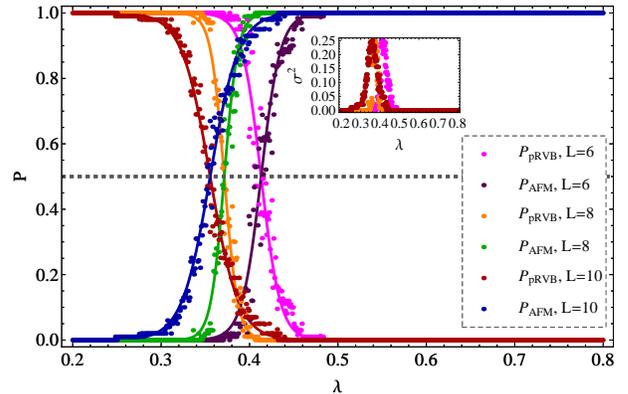}
\caption{\label{Fig:Pdwave} The same as in FIG. \ref{Fig:Pswave} for the frustrated case with the NNN coupling $J_2=J_1$ in the Hamiltonian (\ref{eq:Hamiltonian}). Here the probabilities are for the d-wave pRVB ($P_{pRVB}$) and CAFM phases ($P_{AFM}=1-P_{pRVB}$) and $\lambda_c (L) \approx 0.413$, $0.371$ and $0.355$  for $L=6$, $8$ and $10$, respectively. The probabilities and fits are calculated following the same procedure as in FIG. \ref{Fig:Pswave}.
}
\end{figure}
\begin{figure}[t!]
\includegraphics[width=\columnwidth]{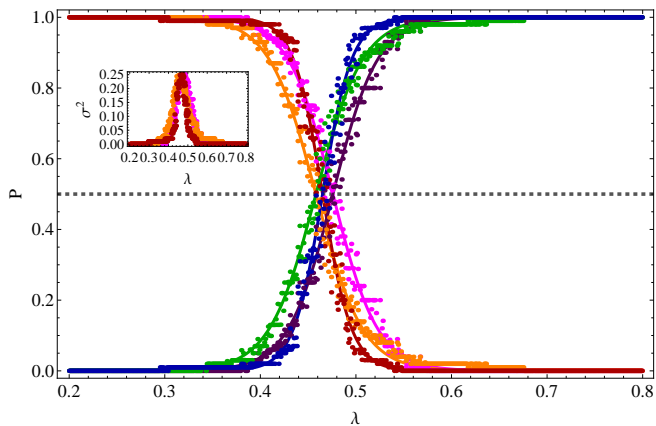}
\caption{\label{Fig:PswaveL10} Probabilities of the s-wave pRVB ($P_{pRVB}$) and NAFM phases ($P_{AFM}=1-P_{pRVB}$) for the unfrustrated case with NNN couplings $ J_2=0$ in the Hamiltonian (\ref{eq:Hamiltonian}) and the system size $N=10\times 10$ lattice sites with periodic boundary conditions for three different CNN architectures. The probabilities and fits are calculated following the same procedure as in FIG. \ref{Fig:Pswave}. All three architectures have one fully connected layer and three convolutional layers. Red and blue data points correspond to a CNN with 256 filters in each convolutional layer and 256 neurons in the fully connected layer, magenta and purple data points to a CNN with convolutional layers containing 64, 128 and 256 filters and one fully connected layer with 256 neurons, and orange and green data points to a CNN with 128 filters in each convolutional layer and 128 neurons in the fully connected layer.  All three network architectures give similar value of the inter-plaquette coupling strength parameter at which the QPT occurs:  $\lambda_c(L)\approx 0.465$, $0.476$ and $0.458$. 
}
\end{figure}

The final layer of a CNN is the output layer where each neuron has softmax activation function that predicts a single class of $K$ mutually exclusive classes and normalizes total probability over predicted output classes to one. In our calculations the number of neurons in the output layer is two corresponding to the number of different phase labels of the input data set. Output probabilities $P_1$ and $P_2$, with $P_1+P_2=1$, correspond to the probabilities of the system to be in AFM or pRVB phase. 

In our SML calculations the CNN consists of three convolutional layers with each layer having 256 filters, one fully connected layer with 256 neurons and an output layer that classifies an input image in one of the two classes corresponding either to pRVB or AFM phase. Calculations are performed for the system sizes of $N=L\times L$ lattice sites with $L=6$, $8$ and $10$ and with periodic boundary conditions where for each $L$ the input spin-spin correlation matrix images are resized to $L\times L$ pixel values (using OpenCV Python machine learning software library resize function). The ground-state spin-spin correlation matrices are calculated with the RBM based RL algorithm  in the inter-plaquette coupling strength parameter $\lambda\in\left[0,0.2\right]$ and $\lambda\in\left[0.8,1.0\right]$ intervals (2000 matrices in each interval) that do not contain the transition point and random 10\% of data separated as a validation data set. The remaining 90$\%$ of data is then used to train the network. The network then predicts class labels for 600 matrices calculated in the interval $\lambda\in\left[0.2,0.8\right]$ and detects the QPT. The probabilities are calculated as averages over hundred CNN output label predictions, each corresponding to one random choice of 10$\%$ of the validation data samples and the network training with the remaining  90$\%$ of data calculated in intervals $\lambda\in\left[0,0.2\right]$ and $\lambda\in\left[0.8,1.0\right]$. Obtained probabilities are shown in FIG. \ref{Fig:Pswave} for the unfrustrated case with the NNN couplings $J_2=0$ and in FIG. \ref{Fig:Pdwave} for the frustrated case with $J_2=J_1$. The QPT point corresponds to the point where the network is maximally confused, that is to the point where $P_{pRVB}=P_{AFM}=0.5$.

To achieve accurate extrapolation of the results to the thermodynamic limit ($L\rightarrow\infty$) and find $\lambda_c(L\rightarrow \infty)$ values it would be necessary to do all presented calculations for much larger system sizes. Particularly difficult and time consuming task would be to generate large enough training data sets for even larger system sizes with described RBM based RL method. However even from the presented results for the system sizes of $N=L\times L$ lattice sites with $L=6$, $8$ and $10$ and periodic boundary conditions it is clearly visible that the obtained values for $\lambda_c(L)$ as the system size is increased rapidly approach the $\lambda_c(L\rightarrow \infty)$ values obtained previously with other methods. In particular the values obtained with the CCM and the real-space RG method are $\lambda_c^{CCM}(L\rightarrow\infty)= 0.47$\cite{Gotze} and $\lambda_c^{RG}(L\rightarrow\infty)=4822$\cite{Fledderjohann} for the unfrustrated case with $J_2=0$, and $\lambda_c^{CCM}=0.301$\cite{Gotze} for the frustrated case with $J_2=J_1$. For the largest system size of $N=10\times 10$ lattice sites with periodic boundary conditions we obtain the values $\lambda_c\approx 0.465$ and $\lambda_c\approx 0.355$ for $J_2=0$ and $J_2=J_1$, respectively, as shown in FIG. \ref{Fig:Pswave} and FIG. \ref{Fig:Pdwave}. 

However, the result of the QMC and finite-size scaling method for the unfrustrated case is $\lambda_c^{QMC}\approx 0.5485$.\cite{Albuquerque,Wenzel,Ran} Our result that is still in reasonable agreement with the QMC result, however indicates slightly larger NAFM region than QMC calculations. To reconfirm our result for the unfrustrated case we perform further calculations with two additional CNN architectures. Both additional network architectures have one fully connected layer and three convolutional layers. The first additional architecture has 128 filters in each convolutional layer and 128 neurons in the fully connected layer, and the second architecture 64, 128 and 256 filters in three convolutional layers and 256 neurons in the fully connected layer. The results for the system size of $N=10\times 10$
lattice sites are shown in FIG. \ref{Fig:PswaveL10}. 

As expected for the initial CNN architecture with 256 filters in each convolutional layer variances of the predictions are smaller than for the other two CNN architectures with reduced number of filters in convolutional layers, since increasing the number of filters improves the accuracy of the predictions resulting in steeper probability lines near the transition. Nonetheless all three architectures give similar values of $\lambda_c(L)$ confirming that over-fitting does not happen in the more complex CNN model training step. Whilst the QMC results suggest that the real space RG results converge non-monotonously towards the value obtained with the QMC calculations when further excited states with more that one quintuplet are included in the RG calculations,\cite{Fledderjohann} our results indicate that the RG results might converge monotonously towards the value close to the value obtained with the CCM.

\section{Conclusions}
\label{sec:Conculsions}
In conclusion we have studied the spin-$1/2$ square lattice $J_1$-$J_2$ model with additional plaquette structure, relevant for the high-temperature superconducting materials and realizable in cold atom experiments, using ML techniques. We have demonstrated the power of these techniques, in particular RBM bases RL algorithm and CNN based SML method, to study complex phases and phase transitions in challenging many-body problems even in the presence of frustration where other important techniques like QMC are limited. Our results show good agreement with previously obtained results, particularly with the CCM and the real-space RG results. 

An interesting direction for the future research would be to explore the power of the presented ML techniques to study dynamical properties of similar complex many-body quantum systems, particularly when the presence of certain dynamical symmetries leads to breaking of the continuous time-translation symmetry and results in existence of a set of observables that can enter robust non-equilibrium limit cycles.

\begin{acknowledgments}
We thank Tomislav \v{S}eva, Hrvoje Buljan, Robert Pezer, and Danko Radi\'{c} for helpful discussions. This work was supported  by the QuantiXLie Centre of Excellence, a project cofinanced by the Croatian Government and European Union through the European Regional Development
Fund - the Competitiveness and Cohesion Operational Programme (Grant KK.01.1.1.01.0004).
\end{acknowledgments}


\begin{thebibliography}{99}
\bibitem{Hershey}J. R. Hershey, S. J. Rennie, P. A. Olsen, and T. T. Kristjansson, Comput. Speech Lang. \textbf{24}, 45-66 (2010).
\bibitem{Silver}D. Silver, A. Huang, C. J. Maddison,  A. Guez, L. Sifre, G. van den Driessche, J. Schrittwieser, I. Antonoglou, V. Panneershelvam, M. Lanctot, S. Dieleman, D. Grewe, J. Nham, N. Kalchbrenner, I. Sutskever, T. Lillicrap, M. Leach, K. Kavukcuoglu, T. Graepel, and D. Hassabis, Nature \textbf{529}, 484-489 (2016).
\bibitem{LeCun}Y. LeCun,  Y. Bengio, and G. Hinton,  Nature \textbf{521},  436 (2015).
\bibitem{Goodfellow}I. Goodfellow, Y. Bengio, A. Courville, and Y. Bengio, \emph{Deep learning}, Vol. 1 (MIT press Cambridge, 2016).
\bibitem{Kelleher}John D. Kelleher, \emph{Deep learning} (MIT press Cambridge, 2019).
\bibitem{Mehta}P. Mehta, and D. J. Schwab, arXiv:1410.3831.
\bibitem{Chung}J.-H. Chung, and Y.-J. Kao, arXiv:2010.05703. 
\bibitem{Koch}E. De Mello Koch, R. De Mello Koch, and L. Cheng, IEEE Access \textbf{8}, 106487-106505 (2020).
\bibitem{Janusz}M. Koch-Janusz, and Z. Ringel, Nat. Phys. \textbf{14}, 578--582 (2018).
\bibitem{Funai}S. S. Funai, and D. Giataganas, Phys. Rev. Research \textbf{2}, 033415 (2020).
\bibitem{Mechta1} P. Mehta, M. Bukov, C.-H. Wang, A. G. R. Day, C. Richardson, C. K. Fisher, and D. J. Schwab, Phys. Rep. \textbf{810}, 1-124 (2019). 
\bibitem{Carleo1}G. Carleo, I. Cirac, K. Cranmer, L. Daudet, M. Schuld, N. Tishby, L. Vogt-Maranto, L. Zdeborov\'{a}, Rev. Mod. Phys. \textbf{91}, 045002 (2019). 
\bibitem{DasSarma}S. Das Sarma, D.-L. Deng, and L.-M. Duan, Phys. Today \textbf{72}(3), 48-54 (2019).
\bibitem{Jia}Z.-A. Jia, B. Yi, R. Zhai, Y.-C. Wu, G.-C. Guo, G.-P. Guo, Adv. Quantum Technol. \textbf{2}, 1800077 (2019). 
\bibitem{Carleo}G. Carleo, and M. Troyer, Science \textbf{355}, 602 (2017). 
\bibitem{Carleo2}G. Carleo, Y. Nomura, and M. Imada, Nat. Commun. \textbf{9}, 5322 (2018). 
\bibitem{Szabo}Attila Szab\'{o} and Claudio Castelnovo, Phys. Rev. Research \textbf{2}, 033075 (2020). 
\bibitem{Shi}H.-Q. Shi, X.-Y. Sun, and D.-F. Zeng, Commun. Theor. Phys. \textbf{71}, 1379 (2019). 
\bibitem{Borin}A. Borin, and D. A. Abanin, Phys. Rev. B \textbf{101}, 195141 (2020). 
\bibitem{Wu}Y. Wu, C. Wei, S. Qin, Q. Wen, F. Gao, arXiv:2005.11970. 
\bibitem{Choo}K. Choo, T. Neupert, and G. Carleo, Phys. Rev. B \textbf{100}, 125124 (2019).
\bibitem{Vieijra}T. Vieijra, C. Casert, J. Nys, W. De Neve, J. Haegeman, J. Ryckebusch, and F. Verstraete, Phys. Rev. Lett. \textbf{124}, 097201 (2020).
\bibitem{Nomura}Y. Nomura, A. S. Darmawan, Y. Yamaji, and M. Imada, Phys. Rev. B \textbf{96}, 205152 (2017).
\bibitem{Nomura2}J. Nomura, arXiv:2009.14777. 
\bibitem{Astrakhantsev}N. Astrakhantsev, T. Westerhout, A. Tiwari, K. Choo, A. Chen, M. H. Fischer, G. Carleo, and T. Neupert, arXiv:2101.08787.
\bibitem{Irikura}N. Irikura, and H. Saito, Phys. Rev. Research \textbf{2}, 013284 (2020). 
\bibitem{Choo2}K. Choo, G. Carleo, N. Regnault, and T. Neupert, Phys. Rev. Lett. \textbf{121}, 167204 (2018). 
\bibitem{Westerhout}T. Westerhout, N. Astrakhantsev, K. S. Tikhonov, M. I. Katsnelson, and A. A. Bagrov, Nat. Commun. \textbf{11}, 1593 (2020). 
\bibitem{Hartmann}M. J. Hartmann, and G. Carleo, Phys. Rev. Lett. \textbf{122}, 250502 (2019). 
\bibitem{Schmitt}M. Schmitt, and M. Heyl, Phys. Rev. Lett. \textbf{125}, 100503 (2020).
\bibitem{Vicentini}F. Vicentini, A. Biella, N. Regnault, and C. Ciuti, Phys. Rev. Lett. \textbf{122} (2019).
\bibitem{Nagy}A. Nagy and V. Savona, Phys. Rev. Lett. \textbf{122} (2019).
\bibitem{Yoshioka}N. Yoshioka, and R. Hamazaki, Phys. Rev. B \textbf{99}, 214306 (2019). 
\bibitem{Liang}X. Liang, W.-Y. Liu, P.-Z. Lin, G.-C. Guo, Y.-S. Zhang, and L. He, Phys. Rev. B \textbf{98}, 104426 (2018). 
\bibitem{Gao}X. Gao, and L.-M. Duan, Nat. Commun. \textbf{8}, 662 (2017). 
\bibitem{Deng}D.-L. Deng, X. Li, and S. Das Sarma, Phys. Rev. X \textbf{7}, 021021 (2017).
\bibitem{Torlai}G. Torlai, G. Mazzola, J. Carrasquilla, M. Troyer, R. Melko, and G. Carleo, Nat. Phys. \textbf{14}, 447-450 (2018). 
\bibitem{Deng2}D.-L. Deng, Phys. Rev. Lett. \textbf{120}, 240402 (2018).
\bibitem{Saito}H. Saito, J. Phys. Soc. Jpn. \textbf{86}, 093001 (2017). 
\bibitem{Sharir}O. Sharir, Y. Levine, N. Wies, G. Carleo, A. Shashua, Phys. Rev. Lett. \textbf{124}, 020503 (2020).
\bibitem{Kaubruegger}R. Kaubruegger, L. Pastori, J. C. Budich, Phys. Rev. B \textbf{97}, 195136 (2018). 
\bibitem{Duric}T. \DJ uri\'c, and T. \v{S}eva, Phys. Rev. B \textbf{102}, 085104 (2020). 
\bibitem{Pilati}S. Pilati, and P. Pieri, Phys. Rev. E \textbf{101}, 063308 (2020). 
\bibitem{Deng3}D.-L. Deng, X. Li, S. Das Sarma, Phys. Rev. B \textbf{96}, 195145 (2017). 
\bibitem{Clark}S. R. Clark, J. Phys. A: Math. Theor. \textbf{51} 135301 (2018). 
\bibitem{Glasser}I. Glasser, N. Pancotti, M. August, I. D. Rodriguez, and J. I. Cirac, Phys. Rev. X \textbf{8}, 011006 (2018). 
\bibitem{Chen}J. Chen, S. Cheng, H. Xie, L. Wang, and T. Xiang, Phys. Rev. B \textbf{97}, 085104 (2018). 
\bibitem{Sutton}R. S. Sutton, and A. G. Barto, \emph{Reinforcement Learning: An Introduction} (MIT press Cambridge, 2018).
\bibitem{Otterlo}M. van Otterlo, and M. Wiering, \emph{Reinforcement learning and Markov decision processes} (Springer-Verlag Berlin Heidelberg, 2012).
\bibitem{Carrasquilla}J. Carrasquilla, Advances in Physics: X, 5:1 (2020).
\bibitem{Carrasquilla2}J. Carrasquilla, and R. G. Melko, Nat. Phys. \textbf{13}, 431-434 (2017). 
\bibitem{Lian}W. Lian, S.-T. Wang, S. Lu, Y. Huang, F. Wang, X. Yuan, W. Zhang, X. Ouyang, X. Wang, X. Huang, L. He, X. Chang, D.-L. Deng, and L.-M. Duan, Phys. Rev. Lett. \textbf{122}, 210503 (2019). 
\bibitem{Iakovlev}I. A. Iakovlev, O. M. Sotnikov, and V. V. Mazurenko, Phys. Rev. B \textbf{98}, 174411 (2018). 
\bibitem{Broecker}P. Broecker, J. Carrasquilla, R. G. Melko, and S. Trebst, Sci. Rep. \textbf{7}, 8823 (2017). 
\bibitem{Dong}X.-Y. Dong, F. Pollmann, X.-F. Zhang, Phys. Rev. B \textbf{99}, 121104 (2019). 
\bibitem{Hsu}Y.-T. Hsu, X. Li, D.-L. Deng, and S. Das Sarma, Phys. Rev. Lett. \textbf{121}, 245701 (2018). 
\bibitem{Hu}W. Hu, R. R. P. Singh, and R. T. Scalettar, Phys. Rev. E \textbf{95}, 062122 (2017). 
\bibitem{Chng}K. Ch'ng, J. Carrasquilla, R. G. Melko, and E. Khatami, Phys. Rev. X \textbf{7}, 031038 (2017). 
\bibitem{Zhang}Y. Zhang, and E.-A. Kim, Phys. Rev. Lett. \textbf{118}, 216401 (2017). 
\bibitem{Wang}L. Wang, Phys. Rev. B \textbf{94}, 195105 (2016). 
\bibitem{Nieuwenburg}E. P. L. van Nieuwenburg, Y.-H. Liu, and S. D. Huber, Nat. Phys. \textbf{13}, 435-439 (2017). 
\bibitem{Broecker2}P. Broecker, F. F. Assaad, and S. Trebst, \\arXiv:1707.00663. 
\bibitem{Yoshioka2}N. Yoshioka, Y. Akagi, and H. Katsura, Phys. Rev. B 97, 205110 (2018). 
\bibitem{Wetzel}S. J. Wetzel, Phys. Rev. E \textbf{96}, 022140 (2017). 
\bibitem{RodriguezNieva}J. F. Rodriguez-Nieva, M. S. Scheurer, Nature Physics \textbf{15}, 790-795 (2019). 
\bibitem{Venderley}J. Venderley, V. Khemani, E.-A. Kim, Phys. Rev. Lett. \textbf{120}, 257204 (2018). 
\bibitem{Giannetti}C. Giannetti, B. Lucini, and D. Vadacchino, Nucl. Phys. B \textbf{944}, 114639 (2019).
\bibitem{Zhang2}W. Zhang, J. Liu, T.-C. Wei, Phys. Rev. E \textbf{99}, 032142 (2019). 
\bibitem{Zhang3}Y. Zhang, R. G. Melko, and E.-A. Kim, Phys. Rev. B \textbf{96}, 245119 (2017).
\bibitem{Berezutskii}A. Berezutskii, M. Beketov, D. Yudin, Z. Zimbor\'{a}s, and J. Biamonte, J. Phys. Complex. \textbf{1} 03LT01 (2020). 
\bibitem{Kharkov} Y. A. Kharkov, V. E. Sotskov, A. A. Karazeev, E. O. Kiktenko, and A. K. Fedorov, Phys. Rev. B \textbf{101}, 064406 (2020).
\bibitem{Rem}B. S. Rem, N. K\"{a}ming, M. Tarnowski, L. Asteria, N. Fl\"{a}schner, C. Becker, K. Sengstock, and C. Weitenberg, Nat. Phys. \textbf{15}, 917-920 (2019).
\bibitem{Rao}W.-J. Rao, J. Phys.: Condens. Matter \textbf{30} 395902 (2018).
\bibitem{Canabarro}A. Canabarro, F. F. Fanchini, A. L. Malvezzi, R. Pereira, and R. Chaves, Phys. Rev. B \textbf{100}, 045129 (2019).
\bibitem{Wetzel2}S. J. Wetzel, and M. Scherzer, Phys. Rev. B \textbf{96}, 184410 (2017). 
\bibitem{Morningstar}A. Morningstar, and R. G. Melko, J. Mach. Learn. Res. \textbf{18}, 5975 (2018).
\bibitem{Zhao1}X. L. Zhao, and L. B. Fu, Ann. Phys. \textbf{410}, 167938 (2019).
\bibitem{Cheng}S. Cheng, F. He, H. Zhang, K.-D. Zhu, and  Y. Shi, arXiv:2101.08928.
\bibitem{Jadrich} R. B. Jadrich, B. A. Lindquist, and T. M. Truskett, J. Chem. Phys. \textbf{149}, 194109 (2018).
\bibitem{Kim}D. Kim, and D.-H. Kim, Phys. Rev. E \textbf{98}, 022138 (2018). 
\bibitem{Bohrdt}A. Bohrdt, C. S. Chiu, G. Ji, M. Xu, D. Greif, M. Greiner, E. Demler, F. Grusdt, and M. Knap, Nat. Phys. \textbf{15}, 921-924 (2019). 
\bibitem{Rzadkowski}W. Rzadkowski, N. Defenu, S. Chiacchiera, A. Trombettoni, and G. Bighin, New J. Phys. \textbf{22} (2020) 093026. 
\bibitem{Yao}J. Yao, Y. Wu, J. Koo, B. Yan, and H. Zhai, Phys. Rev. Research \textbf{2}, 013287 (2020). 
\bibitem{Che}Y. Che, C. Gneiting, T. Liu, and F. Nori, Phys. Rev. B \textbf{102}, 134213 (2020). 
\bibitem{Ohtsuki}T. Ohtsuki, and T. Mano, J. Phys. Soc. Jpn. \textbf{89}, 022001 (2020). 
\bibitem{Zlabys} G. \v{Z}labys, M. Ra\v{c}i\={u}nas, and E. Anisimovas, J. Phys. A: Math. Theor. \textbf{53}, 115302 (2020). 
\bibitem{Tan} D.-R. Tan, C.-D. Li, W.-P. Zhu and F.-J. Jiang, New J. Phys. \textbf{22}, 063016 (2020). 
\bibitem{Lee}P. A. Lee, N. Nagaosa, and X.-G. Wen, Rev. Mod. Phys. \textbf{78}, 17 (2006).
\bibitem{Manousakis}E. Manousakis, Rev. Mod. Phys. \textbf{63}, 1 (1991).
\bibitem{Harland}M. Harland, M. I. Katsnelson, and A. I. Lichtenstein, Phys. Rev. B \textbf{94}, 125133 (2016). 
\bibitem{Altman}E. Altman, and A. Auerbach, Phys. Rev. B \textbf{65}, 104508 (2002). 
\bibitem{Goral}K. G\'oral, L. Santos, and M. Lewenstein, Phys. Rev. Lett. \textbf{88}, 170406 (2002).
\bibitem{Lahaye}T. Lahaye, C. Menotti, L. Santos, M. Lewenstein, and T. Pfau, Rep. Prog. Phys. \textbf{72}, 126401 (2009).
\bibitem{Nascimbene}S. Nascimb\`ene, Y.-A. Chen, M. Atala, M. Aidelsburger, S. Trotzky, B. Paredes, and I. Bloch, Phys. Rev. Lett. \textbf{108}, 205301 (2012). 
\bibitem{Rey}A. M. Rey, R. Sensarma, S. Foelling, M. Greiner, E. Demler, and M.D. Lukin, Europhys. Lett. \textbf{87}, 60001 (2009). 
\bibitem{Olschlager} M. \"{O}lschl\"{a}ger, G. Wirth, T. Kock, and A. Hemmerich, Phys. Rev. Lett. \textbf{108}, 075302 (2012).
\bibitem{Kogeyama}H. Kageyama, T. Kitano, N. Oba, M. Nishi, S. Nagai, K. Hirota, L. Viciu, J. B. Wiley, J. Yasuda, Y. Baba, Y. Ajiro, and K. Yoshimura, J. Phys. Soc. Jpn. \textbf{74}, 1702-1705 (2005).
\bibitem{Ueda}H. T. Ueda, and K. Totsuka, Phys. Rev. B \textbf{76}, 214428 (2007). 
\bibitem{Stock}C. Stock, W. J. L. Buyers, R. A. Cowley, P. S. Clegg, R. Coldea, C. D. Frost, R. Liang, D. Peets, D. Bonn, W. N. Hardy, and R. J. Birgeneau, Phys. Rev. B \textbf{71}, 024522 (2005).
\bibitem{Tranquada}J. Tranquada, H. Woo, T. Perring, H. Goka, G. Gu, G. Xu, M. Fujita, and K. Yamada, J. Phys. Chem. Solids \textbf{67}, 511 (2006). 
\bibitem{Shen}K. M. Shen, F. Ronning, D. H. Lu, F. Baumberger, N. J. C. Ingle, W. S. Lee, W. Meevasana, Y. Kohsaka, M. Azuma, M. Takano, H. Takagi, and Z.-X. Shen, Science \textbf{307}, 901 (2005). 
\bibitem{Ma}N. Ma, P. Weinberg, H. Shao, W. Guo, D.-X. Yao, and A. W. Sandvik, Phys. Rev. Lett. \textbf{121}, 117202 (2018). 
\bibitem{Fritz} L. Fritz, R. L. Doretto, S. Wessel, S. Wenzel, S. Burdin, and M. Vojta, Phys. Rev. B \textbf{83}, 174416 (2011). 
\bibitem{Schmidt}H.-J. Schmidt, J. Phys. A: Math. Gen. \textbf{38}, 2123 (2005). 
\bibitem{Singh}R. R. P. Singh, M. P. Gelfand, and D. A. Huse, Phys. Rev. Lett. \textbf{61}, 2484 (1988). 
\bibitem{Wenzel2}S. Wenzel, L. Bogacz, and W. Janke, Phys. Rev. Lett. \textbf{101}, 127202 (2008). 
\bibitem{Jiang}F.-J. Jiang, Phys. Rev. B \textbf{85}, 014414 (2012). 
\bibitem{Leite}L. S. G. Leite, and R. L. Doretto, Phys. Rev. B \textbf{100}, 045113 (2019).
\bibitem{Merchant}P. Merchant, B. Normand, K. W. Kr\"{a}mer, M. Boehm, D. F. McMorrow, and Ch. R\"{u}egg, Nat. Phys. \textbf{10}, 373-379 (2014). 
\bibitem{Sachdev3}S. Sachdev, \emph{Quantum Phase Transitions}, 2nd ed (Cambridge University Press, Cambridge, England, 2011).
\bibitem{Sachdev}S. Sachdev, Science \textbf{288}, 475 (2000). 
\bibitem{Sachdev2}S. Sachdev, and B. Keimer, Phys. Today \textbf{64}, 2, 29 (2011).
\bibitem{Sentil1}T. Senthil, A. Vishwanath, L. Balents, S. Sachdev, and M. P. A. Fisher, Science \textbf{303}, 1490 (2004). 
\bibitem{Sentil2}T. Senthil, L. Balents, S. Sachdev, A. Vishwanath, and M. P.A. Fisher, Physical Review B \textbf{70}, 144407 (2004). 
\bibitem{Sentil3}T. Senthil, L. Balents, S. Sachdev, A. Vishwanath, and M. P. A. Fisher, J. Phys. Soc. Jpn. \textbf{74}, 1 (2005).
\bibitem{Balents}L. Balents, L. Bartosch, A. Burkov, S. Sachdev, and K. Sengupta, Physical Review B \textbf{71}, 144508 (2005). 
\bibitem{Levin}M. Levin, and T. Senthil, Phys. Rev. B \textbf{70}, 220403 (2004). 
\bibitem{Sandvik1}A. W. Sandvik, Phys. Rev. Lett. \textbf{98}, 227202 (2007). 
\bibitem{Sandvik2}A. W. Sandvik, Phys. Rev. Lett. \textbf{104}, 177201 (2010). 
\bibitem{Shao}H. Shao, W. Guo, and A. W. Sandvik, Science \textbf{352}, 213 (2016). 
\bibitem{Ma2}N. Ma, G.-Y. Sun, Y.-Z. You, C. Xu, A. Vishwanath, A. W. Sandvik, and Z. Y. Meng, Phys. Rev. B \textbf{98}, 174421 (2018). 
\bibitem{Zhao}B. Zhao, J. Takahashi, and A. W. Sandvik, Phys. Rev. Lett. \textbf{125}, 257204 (2020).
\bibitem{You}Y. You, Z. Bi, and M. Pretko, Phys. Rev. Research \textbf{2}, 013162 (2020). 
\bibitem{Shastry}B. S. Shastry and B. Sutherland, Physica B + C \textbf{108}, 1069 (1981).
\bibitem{Lauchli}A. L\"{a}uchli, S. Wessel, and  M. Sigrist, Phys. Rev. B \textbf{66}, 014401 (2002).
\bibitem{Zayed}M. E. Zayed, Ch. R\"{u}egg, J. Larrea J., A. M. L\"{a}uchli, C. Panagopoulos, S. S. Saxena, M. Ellerby, D. F. McMorrow, Th. Str\"{a}ssle, S. Klotz, G. Hamel, R. A. Sadykov, V. Pomjakushin, M. Boehm, M. Jim\'{e}nez-Ruiz, A. Schneidewind, E. Pomjakushina, M. Stingaciu, K. Conder, and H. M. R\o nnow, Nat. Phys. \textbf{13}, 962 (2017). 
\bibitem{Corboz}P. Corboz, and F. Mila, Phys. Rev. B \textbf{87}, 115144 (2013). 
\bibitem{Zhao2}B. Zhao, P. Weinberg, and A. W. Sandvik, Nat. Phys. \textbf{15}, 678-682 (2019).
\bibitem{Carleo3}G. Carleo, K. Choo, D. Hofmann, J. E. T. Smith, T. Westerhout, F. Alet, E. J. Davis, S. Efthymiou, I. Glasser, S.-H. Lin, M. Mauri, G. Mazzola, C. B. Mendl, E. van Nieuwenburg, O. O'Reilly, H. Th\'{e}veniaut, G. Torlai, and A. Wietek, SoftwareX \textbf{10}, 100311 (2019).
\bibitem{Abadi}M. Abadi, P. Barham, J. Chen, Z. Chen, A. Davis, J. Dean, M. Devin, S. Ghemawat, G. Irving, M. Isard, M. Kudlur, J. Levenberg, R. Monga, S. Moore, D. G. Murray, B. Steiner, P. Tucker, V. Vasudevan, P. Warden, M. Wicke, Y. Yu, and X. Zheng, \emph{TensorFlow: Large-Scale Machine Learning on Heterogeneous Distributed Systems} (2015), Software available from tensorflow.org. 
\bibitem{Voigt}A. Voigt, Comp. Phys. Comm. \textbf{146}, 125 (2002). 
\bibitem{Singh2}R. R. P. Singh, Z. Weihong, C. J. Hamer, and J. Oitmaa, Phys. Rev. B \textbf{60}, 7278 (1999). 
\bibitem{Albuquerque}A. F. Albuquerque, M. Troyer, and J. Oitmaa, Phys. Rev. B \textbf{78}, 132402 (2008).
\bibitem{Fledderjohann}A. Fledderjohann, A. Kl\"{u}mper, and K. H. M\"{u}tter, Eur. Phys. J. B \textbf{72}, 559-565 (2009).  
\bibitem{Gotze}O. G\"{o}tze, S.E. Kr\"{u}ger, F. Fleck, J. Schulenburg, and J. Richter, Phys. Rev. B \textbf{85}, 224424 (2012). 
\bibitem{Wenzel}S. Wenzel, and W. Janke, Phys. Rev. B \textbf{79}, 014410 (2009).
\bibitem{Xu}Y. Xu, Z. Xiong, H.-Q. Wu, and D.-X. Yao, Phys. Rev. B \textbf{99}, 085112 (2019). 
\bibitem{Ran}X. Ran, N. Ma, and D.-X. Yao, Phys. Rev. B \textbf{99}, 174434 (2019).
\bibitem{Kruger}S. E. Kr\"{u}ger, J. Richter, J. Schulenburg, D. J. J. Farnell, and R. F. Bishop, Phys. Rev. B \textbf{61}, 14607 (2000). 
\bibitem{Hirsch}J. E. Hirsch, R. L. Sugar, D. J. Scalapino, and R. Blankenbecler, Phys. Rev. B \textbf{26}, 5033 (1982). 
\bibitem{Loh}E. Y. Loh, Jr., J. E. Gubernatis, R. T. Scalettar, S. R. White, D. J. Scalapino, and R. L. Sugar, Phys. Rev. B \textbf{41}, 9301 (1990). 
\bibitem{Marshall} W. Marshall, Proc. R. Soc. London Ser. A, \textbf{232}, 48 (1955). 
\bibitem{Auerbach}A. Auerbach, \emph{Interacting Electrons and Quantum Magnetism} (Springer, Berlin, 1994).
\bibitem{Schollwock}U. Schollw\"ock, Phys. Rev. B \textbf{58}, 8194 (1998). 
\bibitem{Kolgomorov} A. N. Kolmogorov, Doklady Akademii Nauk SSSR \textbf{108}, 179-182 (1961). English transl. Amer. Math. Soc. Transl. (2) \textbf{28} (1963), 55-59.
\bibitem{Hornik1} K. Hornik, M. Stinchcombe, and H. White, Neural Networks \textbf{2}, 359-366 (1989). 
\bibitem{Hornik2} K. Hornik, Neural Networks \textbf{4}, 251-257 (1991).
\bibitem{LeRoux}N. Le Roux, and Y. Bengio, Neural Computation \textbf{20}, 1631-1649 (2008).
\bibitem{Sorella2}S. Sorella, M. Casula, and D. Rocca, J. Chem. Phys. \textbf{127}, 014105 (2007).
\bibitem{Becca}F. Becca and S. Sorella, \emph{Quantum Monte Carlo Approaches for Correlated Systems} (Cambridge University Press, Cambridge, 2017).
\bibitem{Sorella}S. Sorella, and L. Capriotti, Phys. Rev. B \textbf{61}, 2599 (2000).
\bibitem{Sorella1}S. Sorella, Phys. Rev. B \textbf{64}, 024512 (2001).
\bibitem{Casula1} M. Casula and S. Sorella, J. Chem. Phys. \textbf{119}, 6500 (2003).
\bibitem{Casula2}M. Casula, C. Attaccalite, and S. Sorella, J. Chem. Phys. \textbf{121} 7110 (2004).
\bibitem{Wang2}L. Wang, Z.-C. Gu, F. Verstraete, and X.-G. Wen, Phys. Rev. B \textbf{94}, 075143 (2016).
\bibitem{Mezzacapo}F. Mezzacapo, Phys. Rev. B \textbf{86}, 045115 (2012).
\bibitem{Mezzacapo2}F. Mezzacapo, N. Schuch, M. Boninsegni, and J. I. Cirac, New J. Phys. \textbf{11}, 083026 (2009).
\bibitem{Sra}S. Sra, S. Nowozin, S. J. Wright, \emph{Optimization for machine learning} (MIT press Cambridge, 2011).
\bibitem{Kadanoff1}L. P. Kadanoff, \emph{Statistical Physics: Statics, Dynamics and Renormalization} (World Scientific, Singapore, 2000).
\bibitem{Kadanoff2}L. P. Kadanoff, A. Houghton, and M. C. Yalabik, J. Stat. Phys. \textbf{14}, 171-203 (1976).
\bibitem{Efrati} E. Efrati, Z. Wang, A. Kolan, and L. P. Kadanoff, Rev. Mod. Phys. \textbf{86}, 647 (2014).
\bibitem{Iso}S. Iso, S. Shiba, and S. Yokoo, Phys. Rev. E \textbf{97}, 053304 (2018).
\bibitem{Lin}H. W. Lin, M. Tegmark, and D. Rolnick, J. Stat. Phys. \textbf{168}, 1223-1247 (2017).
\end{thebibliography}
\end{document}